# CONSTRUCTING A TRADITIONAL CHINESE MEDICINE DATA WAREHOUSE APPLICATION

**Wing-Kit Sunny Lam**
School of Electrical Engineering and Computer Science,
Science and Engineering Faculty
Queensland University of Technology (QUT)
Brisbane, Australia
Email: wingkit.lam@connect.qut.edu.au

**Tony Sahama**
School of Electrical Engineering and Computer Science,
Science and Engineering Faculty
Queensland University of Technology (QUT)
Brisbane, Australia
Email: t.sahama@qut.edu.au

**Randike Gajanayake**
SMS Management & Technology
Brisbane, Australia
Email: randike.gajanayake@gmail.com

## Abstract

The explosive growth in the development of Traditional Chinese Medicine (TCM) has resulted in the continued increase in clinical and research data. The lack of standardised terminology, flaws in data quality planning and management of TCM informatics are preventing clinical decision-making, drug discovery and education. This paper argues that the introduction of data warehousing technologies to enhance the effectiveness and durability in TCM is paramount. To showcase the role of data warehousing in the improvement of TCM, this paper presents a practical model for data warehousing with detailed explanation, which is based on the structured electronic records, for TCM clinical researches and medical knowledge discovery.

**Keywords**

TCM, Traditional Chinese Medicine, Data Warehouse

## 1 Introduction

With the on-going changes in the living environment of humans, the increase of physical and mental disease has undergone great shifts. Diseases such as immune dysfunction, cancer, environmental pollution disease and other age-related illnesses have increased significantly, and the existing western chemical treatments are not fully meeting the community's need. As a result, the treatment has been changed from simple disease management to a comprehensive treatment paradigm, which includes prevention, health care, medical treatment and rehabilitation (Chou 2003). Traditional Chinese medicine (TCM) is an ancient system of health care from China. It is a unique system to diagnose and cure illnesses and it was developed more than 2000 years ago. It includes herbal and nutritional therapy, restorative physical exercises, meditation, acupuncture and remedial massage. TCM takes a complete approach to understanding normal functions and disease processes, focusing on the prevention of illness as much as it does on the treatment (MedcineNet.Com 2015). As a result of its incredible success rate, TCM has gained respect around the world. According to China Daily (2015), the World Federation of Chinese Medicine Societies (WFCMS) has established an official relationship with the World Health Organisation (WHO), providing technical support to WHO and cooperating with other non-governmental organisations to promote TCM around the world.

As reported by Helmut Kaiser Consultancy (2015), there are more than 3,000 enterprises currently engaging in the manufacturing and processing of TCM. It was reported that TCM had a financial value of AUD$36.8 billion in 2010, and it will continue to rise to AUD$96.2 billion by 2025. With a global market for TCM, it is no surprise that many countries and multinational companies are keen to invest in its research and development. In 2014, the Australian government signed a memorandum of understanding (MOU) with the Chinese government to boost Australia's position in this global TCM market. The partnership between the Beijing University of Chinese Medicine and the University of Western Sydney will develop a research platform and clinical service at the National Institute of Complementary Medicine (NICM). The focus of this research will be the quality, safety and effectiveness of Chinese medicine. Its aim will be to provide a world-class facility of integrated clinical services, education and research facilitates that will serve the Australian people and help to promote Chinese medicine to the world (Australian Trade Commission 2014).

While TCM development is growing exponentially, large amounts of information are being simultaneously published in the Internet. This includes ancient literatures and clinical researches data, and Internet users can obtain these TCM materials easily by performing searches in engines such as Google, Yahoo or Bing®. According to SEOMoz (2015), these web search engines use their own algorithms to evaluate the query and return the results to users, and the results can be ranked and sorted by:

- Keywords and the relevant degree
- Quality and timeliness of web content
- URL and title of website
- Popularity of website

However, Wikipedia (2015) states that most of the web search engines are commercial ventures supported by advertising revenue and thus they allow advertisers to have their listings ranked higher for a fee. If a search is performed on traditional Chinese medicine treatments for the disease Acute Leukaemia, 261,000 results are returned by Google.com whereas Bing®.com returns 2.1 million results (as shown in Figure 1 and 2)

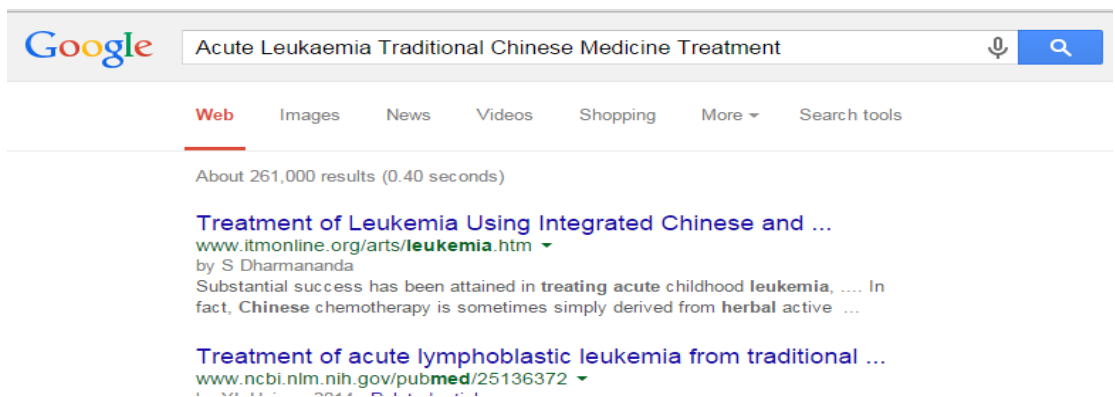

*Figure 1. The search result of "Acute Leukaemia Traditional Chinese Medicine Treatment" in Google.com*

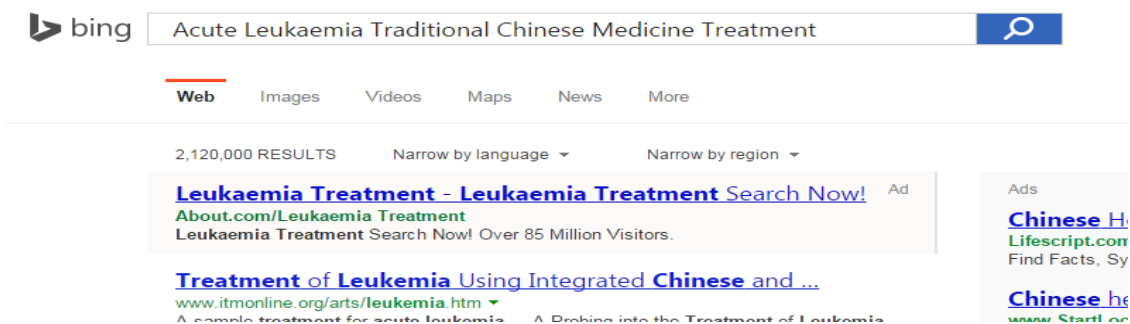

*Figure 2. The search result of "Acute Leukaemia Traditional Chinese Medicine Treatment" in Bing®.com*

With the noticeable difference in the number of results returned by each search engine, obtaining the right information becomes a concern for the users since:

- Excessive or inaccurate information is returned
- Lack analysis methodologies to screen the results

- Large amount of time is required to check and analyse these results

In addition, TCM is commonly stored in the form of written confidential family records. These unstructured documents are dissimilar between communities of practitioners across diverse geographical areas. Consequently there is a lack of standardised terminologies being used and flaws in data quality along with information management become the issues in TCM informatics (Wu & Chen 2008).

Furthermore, clinical trials are the most common clinical research practice in western healthcare industry. It helps improve the diagnosis, treatment and management of people with diseases (NIH 2014). It is able to explore the benefits, detriments and effectiveness of any medical treatment by following strict scientific standards to protect trial participants and the integrity of research. In contrast, TCM knowledge is directly accumulated from daily clinical practices (Zhou 2010). Clinical practice with synthesised treatment based on syndrome differentiation is the core basis of TCM clinical evaluation and clinical study. With the high volume of clinical data available in TCM, the lack of systematic and standardised research processes will hold up TCM development. To support the development of a more complete and cohesive understanding of TCM, adopting wet/dry research methods to examine and analyse this huge amount of clinical data is crucial (Zhou 2010).

In this context, data warehousing is proposed to overcome the data overflow issues in the TCM informatics. There are many reported potential benefits and motivations for data warehouse implementation. In the western healthcare industry, it has been proven that a data warehouse can save time and reduce expenses of organisation. It can also provide health service providers with valuable information to help make better clinical and operational decisions (Caserta Concepts, 2012). According to Inmon (2002), data warehouse is a subject-oriented (can analyse any particular subject area), integrated (integrates data from different data sources), time-variant (keeps historical data) and non-volatile (data kept should never be changed) collection of data to help support management decisions.

The proposed data warehouse system will integrate the diverse TCM data sources and varied data structures. Using OLAP (Online Analytical Processing) and data mining technologies, we can:

- Analyse TCM data in a very short time
- Recommend treatments to health care provider
- Provide more efficient compound formula of treatment
- Discover potential new TCM knowledge

## 2 METHODS

Data warehousing represents the latest standard of database management (Tutorials Point 2015), it includes three technologies:

- Data warehouse - provides generalised and consolidated data in multidimensional views
- Online analytical processing (OLAP) - performs multidimensional analysis of business data and provides the capability for complex calculations, trend analysis, and sophisticated data modelling
- Data mining - integrated with OLAP operations to enhance the interactive mining of knowledge at multiple level of abstraction

Restriction placed on the researcher to obtain a TCM database schema and data from existing database owners to populate the data warehouse, a Redgate Data Generator Tool was used to generate the required datasets to simulate the integration process. All the generated datasets are based on TCM formula information published on the Scared Lotus Chinese Medicine website. Figure 3 illustrates the basic data warehouse workflow.

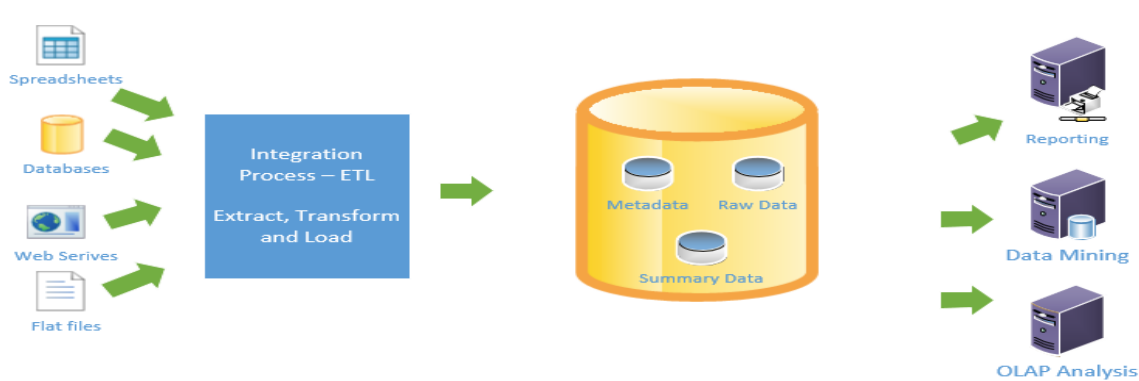

*Figure 3. The basic data warehouse workflow*

A successful data warehouse can provide accurate and timely information to the user. Poolet (2009) has developed the 7D Database Lifecycle Management Method, known as "7D Methods ™". It addresses the lifecycle management of the operational databases and data warehouses, which can be summarised as follows:

- Discover – Investigate business requirement, define the objectives and identify the data sources
- Design – Create conceptual and logical data models for the relational data warehouse and dimensional models to represent the multi-dimensional cubes
- Develop – Create physical data warehouse models, including ETL processes, OLAP cubes, metadata and data dictionary
- Deploy – Deploy data warehouse databases and ETL to production
- Day to Day – Monitor the data warehouse usages, gauge the importance of data and detect when business needs start to shift
- Defend – Create backup and disaster recovery plans
- Decommission – Decommission the data warehouse if it is no longer meets the requirement

In this context, the following steps explain how this "7D Methods ™" methodology are applied on the proposed TCM data warehouse.

## 2.1 Discover – Define Objectives and Identity the data sources

It is import to understand the existing data from the diverse data sources. The implementation of the data warehouse often involves making changes to the database schema. A clear understanding of data relationships among heterogeneous systems is required to determine in advance the impact of any such changes (Syntel 2015). In this paper, the creation of a data warehouse, which can integrate TCM data from disparate data sources, was attempted. OLAP was used to perform multidimensional analysis, while data mining was used to discover the potential of the knowledge and data sharing can effectively promote the exchange of this knowledge. The proposed data warehouse should offer the following advantages:

- Analyse TCM data in a short period of time
- Recommend treatments to health service providers
- Discover the potential of the new TCM knowledge (e.g. more efficient compound formula of treatment)

## 2.2 Design – Create TCM Data Warehouse Model

Data modelling is a design technique for databases intended to support end-user queries in data warehouse. This model should be easily extensible according to future needs. When designing data warehouse model, the most commonly used schema types are Star Schema and Snowflake Schema. These schemas consists of

- Fact table – a table that contains the measures of interest

- Dimension table – a lookup table that provides the details of attributes

According to Oracle (2002), in a star schema design, a single fact table sits in the middle and is radically connected to other surrounding dimension tables (like a star). The primary key in each dimension table is related to a foreign key in the fact table.

A snowflake schema is an extension of the former, the dimensional table is normalized into multiple lookup tables, each representing a level in the dimensional hierarchy. The snowflake schema has been chosen for this research paper due to the nature and the complexity of TCM. The fact table is Formula List and it is surrounded by Formulas, Herbs, Sources, Dates, Formula Types, Herb Types, Countries and Source Types. Figure 4 presents the conceptual model of the proposed TCM data warehouse.

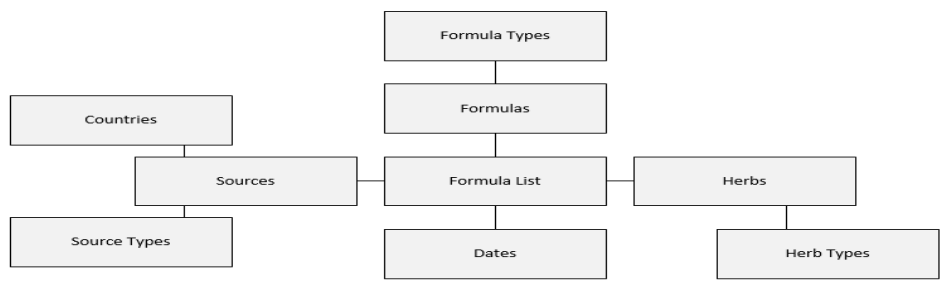

*Figure 4 Snowflake schema diagram for TCM data warehouse*

## 2.3 Develop - Construct TCM Data Warehouse

The data warehouse can be implemented once the conceptual data warehouse model is created. Microsoft SQL Server 2012 (MSQL 2012) has been selected to implement the proposed TCM data warehouse in this paper as it covers all the fundamental technologies to create a data warehouse:

- SQL Server Management Studio - create the physical model of data warehouse
- SQL Server Integration Services - create ETL process
- SQL Server Analysis Services - create cube for OLAP and data mining
- SQL Server Reporting Services - create report by using OLAP queries
- Master Data Services – create Metadata and data dictionary

### 2.3.1 Create physical data warehouse model

SQL Server Management Studio (SSMS) is one of the components of MSQL 2012 that is used to configure, manage and administer all the database components. SSMS was used to create the physical model of TCM data warehouse, including the relational database TCMDW, the fact and dimensional tables as shown in figure 5.

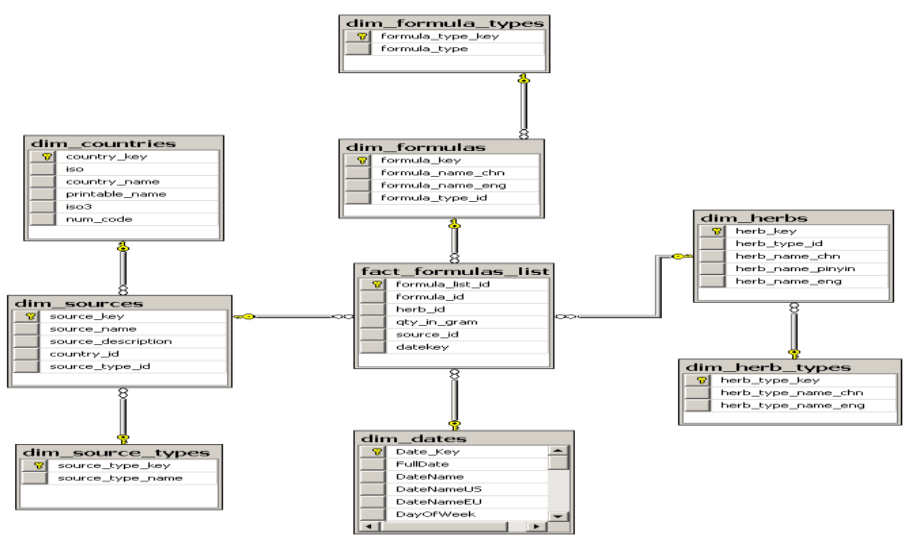

*Figure 5. Physical data warehouse model for TCM data warehouse*

### 2.3.2 Create ETL Processes

The ETL process is responsible for pulling data from all data sources and placing them into the data warehouse. ETL involves the following tasks:

- Extract – is the process to obtain data from different sources

- Transform – is the process of converting the obtained data from its previous format into the proposed data warehouse format. The transformation process occurs by using rules or lookup tables or by combining the data with other data.

- Load – is the process to load the transformed data to the data warehouse

SQL Server Integration Services (SSIS) is a component of MSQL 2012 and it features a fast and flexible data warehousing tool for the ETL process. TCM data can be collected from different sources (e.g. TCM practitioners, hospitals, pharmaceutical companies, research centres, online resources) through different connections (e.g. databases, flat files, web services). By creating individual SSIS packages for each identified data sources, TCM data can be loaded easily from different sources with diverse structures to the proposed data warehouse as shown in figure 6.

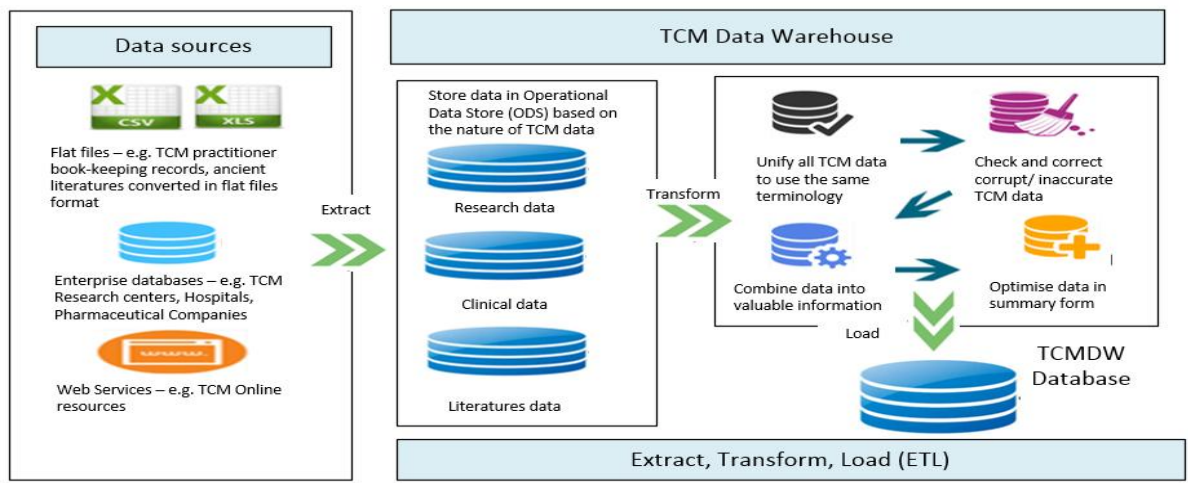

*Figure 6. ETL Process*

### 2.3.3 Create the OLAP Cubes

OLAP cubes are a method of storing data in a multidimensional form generally for reporting purposes (Rouse 2012). It is optimised for the data warehouse and online analytical processing application. OLAP cubes are often pre-summarised across dimensions to drastically improve query response times. According to Thrash (2010), OLAP cubes can provide the following benefits

- Performance - A cube's structure and pre-aggregation allows it to provide very fast responses to queries

- Drill down functionally - Move from summary information to detailed data by focusing in on particular objects.

- Availability of software tools – There are several client software reporting tools that can use an OLAP data source for reporting

SQL Server Analysis Services (SSAS) is an online analytical processing (OLAP), data mining and reporting tool in MSQL 2012. By creating OLAP cubes for the proposed TCM data warehouse, TCM data can be analysed in a very short time and provide more efficient compound formula for treatment.

### 2.3.4 Create Metadata and Data Dictionary

Metadata is crucial to a successful data warehouse implementation. According to Smullins (2010), metadata plays an important role in the data warehouse construction as it describes all the aspects of the data warehouse. Without an understanding of the structure, limitations, definition and description of data, it is likely that data will be misinterpreted or misused. Furthermore if data that is not well defined, it can cause database integrity problems.

To ensure the proposed TCM data warehouse operates in a high quality manner, Master Data Services (MDS) in MSQL 2012 was used to capture data profiles, data modelling, ETL processes, queries and other sources.

### 2.3.5  Generate Reports

Reporting is the presentation layer of data warehouse. SQL Server Reporting Services (SSRS) was used for generating reports in this paper. SSRS is a server-based reporting platform that allows the creation and management of a wide variety of different report types and delivers them in a range of formats. It can create basic reports that contain tables and graphs to more complex data visualisations reports that use charts, maps and spark lines to represent trends.

Examples of these reports will be shown in section 3 of this paper.

## 3   Results

SSRS was used to generate the report below. It compares ingredient consumption for a formula called Ge Gen Tang, a popular treatment for cold and flu. The report presented in Figure 7 shows the average ingredient consumption that prescribed in China comparing with the world in 2010.

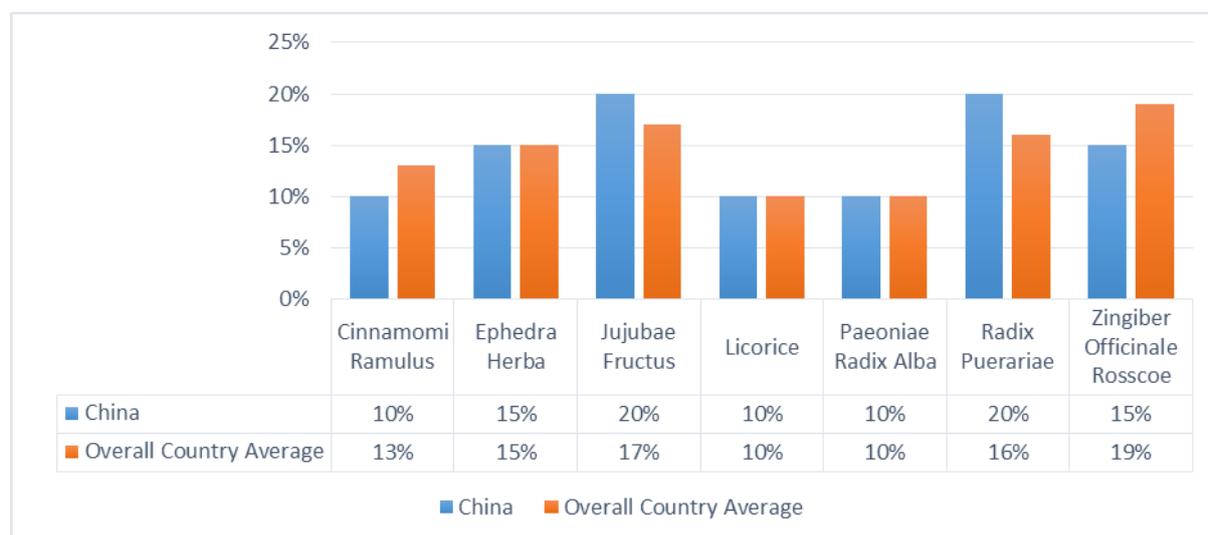

*Figure 7. China versus overall average ingredient consumption of Ge Gen Tang prescribed in 2010*

The results show all the ingredients for the formula that TCM practitioners prescribed in 2010. It can be observed that the base ingredients used were the same at all times, however the ratio of the ingredients was varied depending on the country. Also, a link between region, time and formula ingredient consumption can be detected. According to Health King Enterprise & Balanceuticals Group (2005), one of the principles of TCM in herbal formula is that the formula can be adjusted with addition or reduction of some ingredients to suit the particular need of the patient.

Furthermore, the report data in this paper is based on the data generator tool, the importance of data quality is clearly shown when building a data warehouse as it has direct impact on the accuracy of reports and can lead to incorrect analysis. In the other hand, the well-developed TCM data warehouse can generate analytical and informative reports to enhance the TCM development. For example,

- Diseases trend analysis and prediction
- An unexpected association analysis between diseases, symptoms or formulas
- An unexpected adverse effect analysis of formula or ingredients
- Composite therapeutic approaches analysis

## 4  Discussion

This paper highlighted the benefits of adopting data warehousing technology in TCM development. It demonstrated that this technology is an effective and crucial data management tool for addressing the continued increase in TCM clinical and research data.

We have accomplished the framework and developed the core components in TCM data warehouse design, which includes data modelling, ETL process, OLAP, metadata and reporting functions. The data modelling defines the basic data architecture of TCM data warehouse.  The ETL process transforms TCM data from numerous source systems into a common format. This process assists improvement of TCM data quality and consistency.  OLAP technology provides the capability to inspect the ingredients of a TCM prescribed formula through the multi-dimensional model, which helps to provide more efficient compound formula of treatment and recommend treatments to health service providers. The query execution times are also intensely reduced.  Western pharmaceutical industry realised and applied the of data mining techniques to fully exploit the potential competitive advantage in discovery, development and marketing of their products early in this century (Robert 2001).  Likewise, analysis of TCM data can be performed to identify unknown associations between various ingredients to discover potential new TCM knowledge with these techniques. What is more, the non-volatile and time-variant nature of data warehouses provides historical intelligence, it stores large amounts of historical data and can enable advanced business intelligence including time-period analysis, trend analysis, and trend prediction.

Even though applying data warehouse methodologies can be a key success factor in furthering the TCM development, there are several challenges identified during the life cycle of this project.  One of the major challenges is obtaining database schemas and data from existing TCM owners.  This reluctance prohibited the development of a more detailed framework to implement the TCM data warehouse.  Furthermore, most of the TCM online resources are implemented in the Chinese language. Therefore, it is a laborious job to collect and transform data in high quality structure format.

## 5  Conclusion and Future Work

For demonstration purposes, the scope of this project only considered the Chinese herbal and nutritional therapy.  As stated in the introduction section, the complete TCM data warehouse should also include restorative physical exercises, meditation, acupuncture, and remedial massage.  Also, TCM clinical data are the essential requirements to build the proposed data warehouse.  Obtaining clinical data from TCM hospitals helps in analysing the effectiveness of various medical practices. However, the privacy and security issues are the main concerns in sharing clinical data, information about practitioners and patients must be well protected during ETL process, and this must be addressed in the design phase.  Furthermore, connecting the proposed data warehouse platform with modern western healthcare systems and obtain biochemical information, it can further enhance the possibility to discover more TCM knowledge and advance its development.

## 6  References

Australian Trade Commission. 2014. "Australia Positions Itself for Traditional Chinese Medicine Market" http://www.austrade.gov.au/invest/investor-updates/2014/australia-positions-itself-for-traditional-chinese-medicine-marke Retrieved 28 May, 2015.

Benson, J. 2010. "Traditional Chinese Medicine Sweeps South America" http://www.naturalnews.com/030162_Brazil_Traditional_Chinese_Medicine.html Retrieved 25 February, 2015.

Caserta Concepts. 2012. "Healthcare industry set to benefit from data warehousing and BI use" http://casertaconcepts.com/2012/05/24/data-warehouse-design/ Retrieved 1 March, 2015.

Cawley, K. 2014. "The Benefits of Having a Data Warehouse" http:// cloudtweaks.com/2014/08/benefits-data-warehouse/ Retrieved 1 March, 2015.

Chen, H., Wu, Z. 2007. "Semantic Grid: Model, Methodology, and Applications", pp 195-199. DOI: http:// https://dx.doi.org/10.1007/978-3-540-79454-7 Retrieved 25 April, 2015.

China Daily. 2007. "Chinese Medicine Gets WHO Recognition."

http://www.china.org.cn/china/2015-03/13/content_35039629.htm Retrieved 19 March, 2015.

Chou, W. 2003. "International Natural Medicine Market Analysis", pp147-151
http://lib.cnki.net/cpfd/ZZYY200300001063.html Retrieved 25 February, 2015.

Edelstein, E., and Small, R. 2001. "Data Mining in the Pharmaceutical Industry"
http://www.ddw-online.com/fall-2001/p148598-data-mining-in-the-pharmaceutical-industry.html Retrieved 10 October, 2015.

Health King Enterprise and Balanceuticals Group, Inc. 2005. "Basics of Traditional Chinese Medicine"
http:// www.healthkingenterprise.com/v2/articles/tcm_basic.asp Retrieved 30 May, 2015.

Helmut Kaiser Consultancy. 2015. "New market study, market report, market research, market analysis, market forecast: Traditional Chinese Medicine (TCM) In China and Worldwide 2014-2015-2016-2017-2018-2019- 2020-2025 with History 2012-2013."
http://www.hkc22.com/ChineseMedicine.html Retrieved 28 May, 2015.

HIMSS. 2013. "Clinical & Business Intelligence: Data Management – A foundation for Analytics"
http://www.himss.org/files/himssorg/content/files/201304_data_integration_final.pdf Retrieved 1 March, 2015.

Inmon W. 2002. "Building the Data Warehouse Third Edition", pp31-35
http://citeseerx.ist.psu.edu/viewdoc/download?doi=10.1.1.91.6134&rep=rep1&type=pdf Retrieved 1 March, 2015.

MedicineNet.Com. 2015. "Definition of Oriental Medicine (Traditional)"
http://www.medicinenet.com/script/main/art.asp?articlekey=10786 Retrieved 1 March, 2015.

National Institute of Health (NIH) 2014. "What Are Clinical Trials?"
http://www.nhlbi.nih.gov/studies/clinicaltrials Retrieved 1 October, 2015.

Oracle Corporation. 2002. "Data Warehousing Concept". Oracle9i Data Warehousing Guide Release 2(9.2).
http://docs.oracle.com/cd/B10500_01/server.920/a96520/concept.htm Retrieved 1 March, 2015.

Poolet, M. 2009. "Seven Steps Successful Data Warehouse"
http://sqlmag.com/database-administration/seven-steps-successful-data-warehouse-projects Retrieved 21 April, 2015.

Rouse, M. 2012. "OLAP Cube"
http://searchdatamanagement.techtarget.com/definition/OLAP-cube Retrieved 28 April, 2015.

Sacred Lotus Chinese Medicine. 2015. "TCM Basic. Chinese Formula"
http://www.sacredlotus.com/go/chinese-formulas Retrieved 14 April, 2015.

SEOmoz, Inc. 2007. "How Search Engines Operate"
https://moz.com/beginners-guide-to-seo/how-search-engines-operate Retrieved 14 April, 2015.

Smullins, C. 2010. "On the Importance of Metadata"
http://datatechnologytoday.wordpress.com/2010/09/07/on-the-importance-of-metadata Retrieved 28 April, 2015.

Syntel Inc. 2015. "Eleven Steps to Success in Data Warehousing"
http://www.syntelinc.com/sites/default/files/syntel_dw_11steps.pdf Retrieved 21 April, 2015.

Thrash, W. 2010. "When to use an OLAP cube?"
http://willthrash.com/2010/04/16/when-to-use-an-olap-cube Retrieved 30 April, 2015.

Tutorials Point. 2015. "Data Warehouse Overview"
http:// www.tutorialspoint.com/dwh/dwh_overview.htm Retrieved 21 April, 2015.

Wikipedia. 2015. "Web Search Engine"
http:// en.wikipedia.org/wiki/Web_search_engine Retrieved 1 May, 2015.

Zhou, Xuezhong, et al. 2010. "Building Clinical Data Warehouse for Traditional Chinese Medicine Knowledge Discovery"

http://www.researchgate.net/publication/4347555_Building_Clinical_Data_Warehouse_for_Traditional_Chinese_Medicine_Knowledge_Discovery Retrieved 21 April, 2015.

# Copyright